\documentclass{article}

\usepackage{arxiv}

\usepackage[utf8]{inputenc} % allow utf-8 input
\usepackage[T1]{fontenc}    % use 8-bit T1 fonts
\usepackage{hyperref}       % hyperlinks
\usepackage{url}            % simple URL typesetting
\usepackage{booktabs}       % professional-quality tables
\usepackage{amsfonts}       % blackboard math symbols
\usepackage{nicefrac}       % compact symbols for 1/2, etc.
\usepackage{microtype}      % microtypography
\usepackage{lipsum}		% Can be removed after putting your text content
\usepackage{graphicx}
\usepackage{doi}
\usepackage{subcaption}
\usepackage{float}
\usepackage{caption}
\usepackage{natbib}
\usepackage{tabularx, multirow}
\usepackage[final]{pdfpages}
\usepackage{authblk}

\title{Unsilencing Colonial Archives via Automated Entity Recognition}

\author[]{Mrinalini Luthra}
\author[]{
	Konstantin Todorov}
\author[]{
	Charles Jeurgens}
\author[]{
	Giovanni Colavizza}
\affil[]{\textit{Universiteit van Amsterdam, The Netherlands}}

\date{}

\renewcommand{\headeright}{ }
\renewcommand{\undertitle}{ }
\renewcommand{\shorttitle}{Unsilencing Colonial Archives}

\begin{document}
\maketitle

\begin{abstract}
	Colonial archives are at the center of increased interest from a variety of perspectives, as they contain traces of historically marginalized people. Unfortunately, like most archives, they remain difficult to access due to significant persisting barriers. We focus here on one of them: the biases to be found in historical findings aids, such as indexes of person names, which remain in use to this day. In colonial archives, indexes can perpetuate silences by omitting to include mentions of historically marginalized persons. In order to overcome such limitations and pluralize the scope of existing finding aids, we propose using automated entity recognition. To this end, we contribute a fit-for-purpose annotation typology and apply it on the colonial archive of the Dutch East India Company (VOC). We release a corpus of nearly 70,000 annotations as a shared task, for which we provide baselines using state-of-the-art neural network models. Our work intends to stimulate further contributions in the direction of broadening access to (colonial) archives, integrating automation as a possible means to this end.
\end{abstract}

% keywords can be removed

\section{Introduction}

While archives serve as important resources for researching the past, we must also recognize that they are flawed building blocks for understanding it. As Verne Harris aptly put it, archives only offer “a sliver of a sliver of a sliver”~\citep{harris2002archival}. These snatches of the past usually hide more than contemporary researchers would like to see. The focus of this paper concerns records created in a colonial context, at a time when slavery was rampant. These records mirror the 17\textsuperscript{th} and 18\textsuperscript{th} century hierarchies in terms of religion, race, gender and class and echo the voices and views of the creators of the records, the colonial administrators, the traders, and slave owners. They were the ones who made observations and determined what was worth noting. They did this ---how could it be otherwise--- from a European point of view, in their own interest and not always with a thorough knowledge of the local situation, let alone of the indigenous or enslaved people they encountered. The archives that have been created in this way are therefore not so much a faithful representation of the reality of what went on in the colonial areas but a constructed and selective image of the social order \citep{spivak1985rani,jeurgens2020paradoxes}.

In recent years, the colonial archive has attracted a great deal of scholarly interest; not only from the perspective of archives as a source for historical research but above all as an object of study in itself \citep{ballantyne2004archives,burton2003dwelling,hamilton2002refiguring,bastian2003owning,stoler2010along,lowry2017displaced,namhila2017little}.
This transformation of research interest has led to numerous publications in which power aspects of archiving and the archive are addressed ~\citep{schwartz2002archives, jimerson2009archives, stoler2010along} and has resulted in a shift in perception whereby archives are no longer seen as neutral repositories of sources but as contested spaces of knowledge production. It has also led to more critical reflection on the role archivists play in how they describe and contextualize these records in terms of perpetuating or adjusting one-sided perspectives and power imbalances~\citep{caswell2016human, ghaddar2019go}. Scholars who take the archive as the object of research are not only interested in what archives contain, but as much in what archives do not reveal. After all, the archive “conceals, distorts and silences as much as it reveals”~\citep[48]{fuentes2016dispossessed}. David Thomas asserts that archival silences have now become a proper subject for research and are no longer seen as just passive and annoying gaps in our knowledge \citep{thomas2017silence}. Archival silences are of many types. It was Michelle Rolph Trouillot's~\citep{trouillot2015silencing} groundbreaking book \textit{Silencing the Past}, in which he dissected mechanisms of silencing in the production of history through an analysis of the Haitian revolution and the attention this revolution received in historiography, which gave an impulse to investigate the mechanisms of silencing in the archives. Trouillot distinguished four moments in which silences could enter the process of historical production: 1) the moment of fact creation (which refers to the recording of information: who decides what is recorded, which words are used etc.); 2) the moment of fact assembly (which refers to which recorded information is admitted to become part of the archives); 3) the moment of fact retrieval (which refers to what information is used from the archives to compile a narrative); and 4) the moment of what he called retrospective significance, or the making of history. What interests us most here is the role archivists and archival institutions have played in creating or perpetuating silences. Archivists not only determine to a significant extent which records are admitted in the archives, but also how retrieval of information from the documents is facilitated via archival descriptions, indexes and other technologies designed and used to expedite access to the records from the past. In this paper our focus is on the latter.

If we look at how archivists have facilitated access to the contents of the records, we may conclude that their work was always focused on a very formal way of describing archives. Not the content of documents, but the content of the archive in terms of document types and the function of these documents from the perspective of the formal archive creator were the points of reference for making descriptions. Archivists have for long presented this method as neutral and impartial. The dominant methods of description, Verne Harris and Wendy Duff argued, ``facilitate the needs of certain types of users, but give short to others''~\citep[279]{duff2002stories}. In some cases, archivists chose to improve access to the contents of the documents, for instance by generating indexes on the names mentioned in the documents. To make the labor-intensive indexing manageable, choices were made, for example by indexing only the names of certain categories of persons that appeared in the documents. Those choices mirror what was considered important and what was less important, thus reinforcing the mechanisms of privilege and marginalization by spotlighting certain groups and obscuring others. Only recently these problematic mechanisms have started to receive fundamental attention in the archival discipline~\citep{yeo2017continuing}.

The application of machine learning methods to improve the means of accessing historical records is a growing trend~\citep{colavizza_archives_2022}. A key task in information extraction from texts is that of \textit{named entity recognition and classification}: the automatic ``detection of named entities, i.e., elements in texts which act as a rigid designator for a referent, and their categorisation according to a set of predefined semantic categories''~\citep{ehrmann_named_2021}. Canonical examples include persons, places and organizations. Named entity tasks underpin the automatic creation and enrichment of information systems, allowing to search and browse an archive using as anchors the entities and relations among them established in the very contents of the records. What is more, even indirect (e.g., unnamed) references to entities can be detected, offering a means to surface previously ignored traces. This approach constitutes a radical expansion of the traditional means of access to archival records, as previously described: it allows to navigate the archives more fluidly, across individual records and groups~\citep{ranade_2016}, all the while not discarding archival context to follow the allure of full-text search~\citep{winters_2019}. \textit{Content-based indexing offers new possibilities in surfacing and foregrounding mentions of people who are at present hidden in the documents and historical access tools.}

In this work, our goals are two-fold:
\begin{enumerate}
    \item in general, to propose using automated entity recognition on historical archival records as a means to expand and pluralize access to archives;
    \item specifically, to advance the application of entity recognition to colonial archives, where its use seems most urgent in order to complement critically lacking access tools.
\end{enumerate}

Our contributions include a typology of entities and their attributes designed for automated entity recognition in colonial archives and further adapted to our case study: the Dutch East India Company testament records. We release a shared task in the form of a high-quality corpus of 68,429 annotations. Furthermore, we establish baselines relying on modern machine learning methods. Lastly, our work intends to call for further contributions in the direction of broadening access to colonial archives, taking the form of models, annotations, applications, and their assessment.

\subsection*{Ethical Considerations}

Before turning to the work of archivists in their efforts to provide access to archives, we would like to briefly reflect on one of the most fundamental problems of colonial archives and archives of slavery: the deep and unbridgeable chasm between what these archives offer and the needs of contemporary users seeking traces of their history. In most cases, searching these records by the descendants of colonized and/or enslaved people in the hope of detecting some direct voices from their ancestors leads to disappointment, frustration, and anger after being confronted with the violent categorizations from the past in which enslaved people are described as saleable commodities which turned them into “nonpersons” \citep{hartman2008venus, patterson2018slavery, fuentes2016dispossessed, zijlstra2021voormoeders}.   
There is no doubt that the colonial archive is inherently problematic for those who want to hear snatches of the voices of the enslaved and colonized population. The fact that the colonized and enslaved have left hardly any traces in the colonial archive produced by themselves, does not mean that they do not feature frequently in the archive. The archive is full of records about them. But even then, it is not easy to gain access to these indirect traces of their presence. Feminist historian Durba Ghosh called on historians to also pay attention to the history of the people who have been made nameless and blames them for not looking critically enough at different naming practices since “incompletely named and renamed subjects have histories that are waiting to be told”~\citep[316]{ghosh2004decoding}. She argues that the colonial archives are uneven in their erasure and although the renaming of enslaved people may have led to a sort of archival death, because it is not possible anymore to know where they came from, this does not release researchers from the obligation to ask how they may “decipher women from the archives when they are unnamed?”~\citep[299]{ghosh2004decoding}. She mentions several examples of individuals who can be followed in their life course including the changes in social status after they were given a Christian name. Historian Marjoleine~\cite{kars2020blood} has shown in her recently published book on the massive slave revolt in the Dutch colony of Berbice (which just like the Haitian slave revolt, has received scant attention in historiography) how valuable the indirect traces of renamed enslaved can be. She used 900 testimonies of enslaved persons which were recorded in the interrogation reports after the revolt, which started in 1763 and lasted for one year, was put down. Kars argues that even though the testimonies have been given under pressure, they do provide information about the personal experiences of the enslaved. 

\section{Related Work} 

Our research ties in with two major debates and developments in the humanities: the use of digital methods and artificial intelligence to make extensive historical collections more accessible and, under the heading of `decolonization of the archives', dismantling coloniality in archival infrastructures, for example by developing inclusive and people-oriented search infrastructures.

\subsection{Named Entity Recognition in Historical Documents}

The digitization and extraction of information from existing, historical indexing tools can serve to bootstrap the creation of searchable archival information systems~\citep{colavizza_index-driven_2019,koolen_modelling_2020}. Nevertheless, it is by considering information extraction from the full contents of archival records that a broader, more systematic pluralization of access can be achieved. This is becoming a possibility largely thanks to growing efforts in digitization~\citep{terras_rise_2011} and progress in the automatic extraction of text from handwritten records~\citep{muehlberger_transforming_2019}. \textit{Named entity recognition and classification} (NER for short) constitutes a key step in information extraction pipelines~\citep{ehrmann_named_2021}, and rests at the core of our approach here. In recent years, this task has seen rapid improvements thanks to neural networks~\citep{lample-etal-2016-neural} and pre-trained language models such as BERT~\citep{devlin-etal-2019-bert}. NER is also related, and typically followed up by the tasks of \textit{disambiguation or linking} an entity mention to a knowledge base, and of \textit{relation extraction} among named entities.

The task of named entity recognition and classification on historical documents poses a distinct set of challenges~\citep{ehrmann_named_2021}: document type and domain variety, dynamics of language, noisy input, and lack of resources. While common to the broader application of natural language processing to historical texts~\citep{piotrowski_natural_2012}, these challenges emerge in full force in the context of historical archives. The variety of record typologies (which entail different formats, layouts, and diplomatic characteristics) and the domains of their contents is broad, and only mitigated by a relative uniformity within homogeneous record groups. Language variety and the dynamic change of language use over time are also related issues. All of this has direct implications for the named entities to be extracted, and their typologies. Noise in the inputs is primarily, yet not exclusively, due to the preceding task of automatic text recognition. Such noise can have a significant impact on
downstream tasks, including named entity recognition~\citep{chiron_impact_2017,hill_quantifying_2019,van_strien_assessing_2020}, and can only partially be mitigated by post-hoc correction~\citep{rigaud_icdar_2019,nguyen_survey_2021}. Lastly, NER is a resource-intense task, specifically requiring named entity typologies, lexicons, corpora and more recently pre-trained language embeddings~\citep{ehrmann_named_2016}. All these often lack in a historical setting, or are not immediately re-usable due to the previously mentioned challenges.

Named entity recognition on historical documents is dominated by neural network-based approaches~\citep{ehrmann_named_2021}. Architectures relying on strong, pre-trained embeddings and BiLSTM-CRF layers~\citep{todorov_transfer_2020} or transformer-based models~\citep{boros_robust_2020} typically outperform alternatives. Perhaps as expected, the availability of larger corpora, lower degrees of noise and pre-cleaning texts via heuristics all contribute towards achieving better results~\citep{arampatzis_overview_2020}. What is more, embedding models working at the sub-word level (e.g., character or sequences of characters) mitigate issues due to noise and out of vocabulary forms. Fine-tuning large pre-trained embeddings also helps in general~\citep{gururangan_dont_2020}, and specifically for historical texts~\citep{konle_domain_2020}. Related to our work,~\cite{hendriks_recognising_2020} focus on person NER and entity linkage from notarial archives in Dutch, considering a similar period to ours and records which, at least in part, mention VOC sailors. They start by applying the spaCy~\footnote{\url{https://spacy.io}.} and BERTje~\citep{vries_bertje_2019} models, yielding limited out-of-the-box results. Further fine-tuning spaCy allows them to achieve a maximum precision, recall and F1-score of 0.731, 0.756, 0.743 respectively, in their best fold using fuzzy matching evaluation of at least 90\% between recognized and ground truth mention spans. Their results provide us with an indicative, albeit not fully compatible, comparison in what follows.

\subsection{VOC Testaments}

The establishment of the Dutch United East India Company (Verenigde Oost-Indische Compagnie, VOC), founded in 1602 as a merger of six small local companies, marked the beginning of Dutch expansion in Asia. In 1619, Batavia was founded after the conquest of the Javanese town Jacatra and became the administrative center of Dutch trade and power in Asia. The supreme authority was in the hands of the Governor General and Council of the Indies and in the areas which were under immediate control of the company, local administrative bodies were set up which fell under the authority of the Governor General and Council. 
The VOC settlements, such as Batavia, were complex microcosms, controlled by a very small group of Europeans (in 1680, about 7 percent of the population in Batavia was European), far away from the Dutch Republic, in a society that consisted of many different, mainly Asian populations. In the multi-ethnic and multi-religious settlements, the policy of the VOC was to maintain the status quo by keeping the different groups separate from each other: the free separated from the unfree, Christians separated from non-Christians \citep[197]{brandon2020slavernij}. Christian slave owners were not allowed to resell their slaves to non-Christians and within the VOC logic, Christian slaves were more likely to be released than non-Christians. Some local institutions that were established reflect this division. The ``College van Weesmeesters" [Board of Governors of the Orphan Chamber] was an important facility for Europeans, as it ensured that the estates of deceased Europeans were handled. The counterpart of this board was the ``College van Boedelmeesters van Chinese en andere Onchristen Sterfhuizen'' [Trustees for the Deceased estates of Chinese and other non-Christian Bereaved].
Notaries, although not employed by the Company, played an important role in the larger VOC settlements. Between 1620 and 1650 there were always two public notaries in the city of Batavia and after that three or four. Notaries were obliged to offer their services to everyone, so not only to the Europeans, but also to the Asian population and even to the enslaved people who made up about half of the population of Batavia in the 17\textsuperscript{th} and 18\textsuperscript{th} century. The notaries thereby documented the lives of ordinary people and most of the individuals appearing in these archives belong to the Asian underclass \citep[201-208]{brandon2020slavernij}.
In his book \textit{Batavia}, \cite{niemeijer2012batavia} gives many details of the lives of enslaved people, thus showing the richness of these notarial archives that offer the opportunity to get very close to the local society and its inhabitants. The Indonesian National Archives holds the archives of 111 notaries who resided in Batavia between 1620 and early 19\textsuperscript{th} century, and some archives from notaries who worked in other settlements. These notary archives consist of almost 9000 inventory numbers with documents, deeds and registers with a total size of approximately 1160 linear metres.

\begin{figure}[h!]
    \centering
    \begin{subfigure}[b]{0.45\textwidth}
        \includegraphics[width=0.9\textwidth]{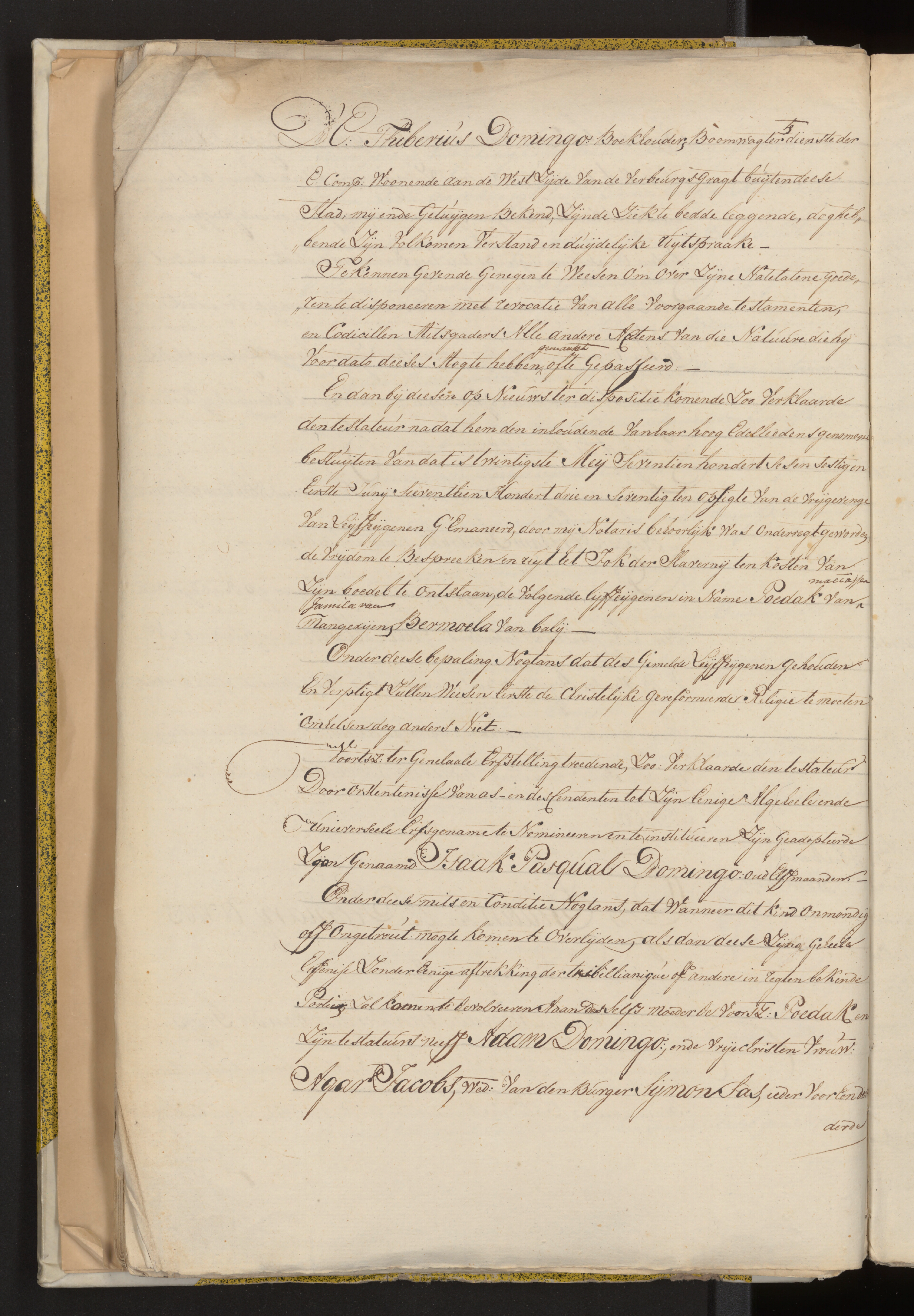}
        \caption{Digital scan of a VOC testament page.}
        \label{fig:scan}
    \end{subfigure}
    \begin{subfigure}[b]{0.45\textwidth}
        \includegraphics[width=0.9\textwidth]{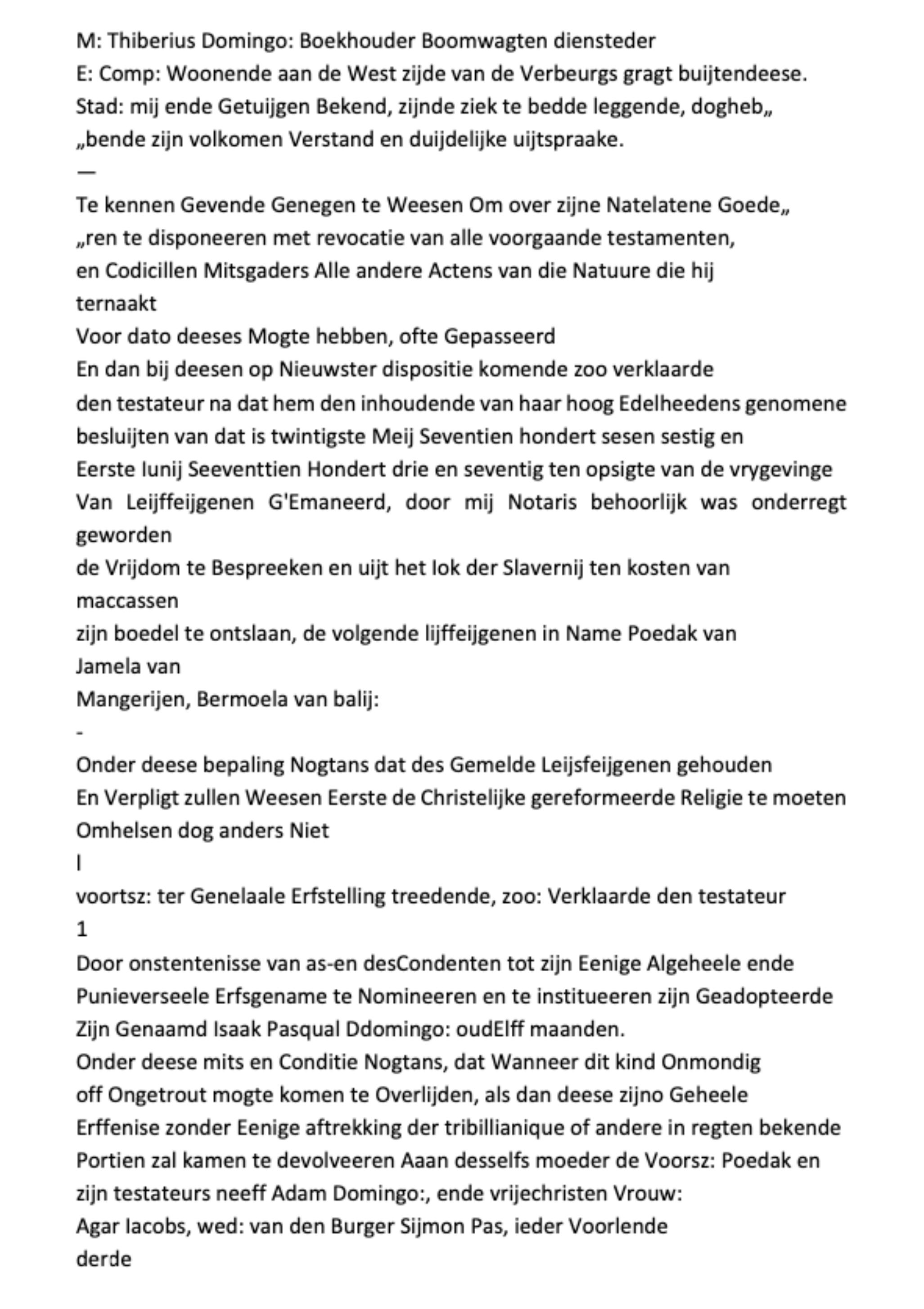}
        \caption{HTR output from Transkribus.}
        \label{fig:htr}
    \end{subfigure}
    \caption{Example of VOC testaments and their HTR processing.}
    \label{fig:Scan/HTR}
\end{figure}

A specific genre of notarial deeds are the testaments (or wills). Due to the high risk of death during their stay in Asia, personnel sent out by the VOC had to draw up a will. The number of personnel employed by the VOC was approximately 11,000 in the 17\textsuperscript{th} century but rose to more than 25,000 in the company’s heyday. Testaments facilitated the settlement of estates after one’s death which was also in the interest of the company. Such a will could be drawn up in various ways: in the Republic, on the VOC ships and in the VOC settlements. Copies of wills made up in the settlements were sent to the VOC headquarters in the Republic after an employee passed away. A relatively small number of 10,000 of these wills (mainly from the 18\textsuperscript{th} century) compiled in 51 bundles (‘banden’ in Dutch) consisting of a total of 53,370 pages survived and are nowadays in custody of the Dutch National Archives. A much larger set of copies of wills are kept in the archive of the Orphan Chamber [Weeskamer] in Batavia in custody of the National Archives in Jakarta.  
Wills drawn up by notaries are an example of a document type that has relatively unambiguous and well-structured diplomatic features. There is a fixed order in the text structure and wills have recurring elements such as the name of the notary, the name or names of the testator(s), the name or names of the beneficiaries (persons or organizations) and the names of witnesses. Dutch archivists compiled a name index of the 10,000 testators of these wills in the nineteenth century and even though in many cases married couples had a will drawn up together, only the names of the European male testators are indexed. It is a clear example of how the work of archivists contributes to silencing already marginalized people. The names of the female co-testators have been left out of the index, as have all the other names appearing in the wills: those of male and female locals of different ethnic backgrounds and the enslaved who appear in many different roles in the wills such as beneficiary, housemate, concubine, creditor, debtor, or property \citep{jeurgens2020paradoxes}. These wills have recently been digitized and made machine-readable with the HTR software\textit{ Transkribus}~\citep{kahle2017transkribus} as part of the Dutch National Archive's\footnote{\url{https://www.nationaalarchief.nl/en}.} digitization project: \textit{IJsberg Zichtbaar Maken}\footnote{\url{https://noord-hollandsarchief.nl/ontdekken/nhalab/ijsberg-zichtbaar-maken}.} [Making the Iceberg Visible]. An example of a scan of a VOC testament\footnote{\url{https://www.nationaalarchief.nl/onderzoeken/archief/1.04.02/invnr/6848/file/NL-HaNA_1.04.02_6848_0150}.} and its corresponding HTR output is provided in Figure \ref{fig:Scan/HTR}. 

\section{Annotation Typology}

\begin{figure}[h!]
    \centering
    \captionsetup{justification=centering}
    \includegraphics[width=.95\linewidth]{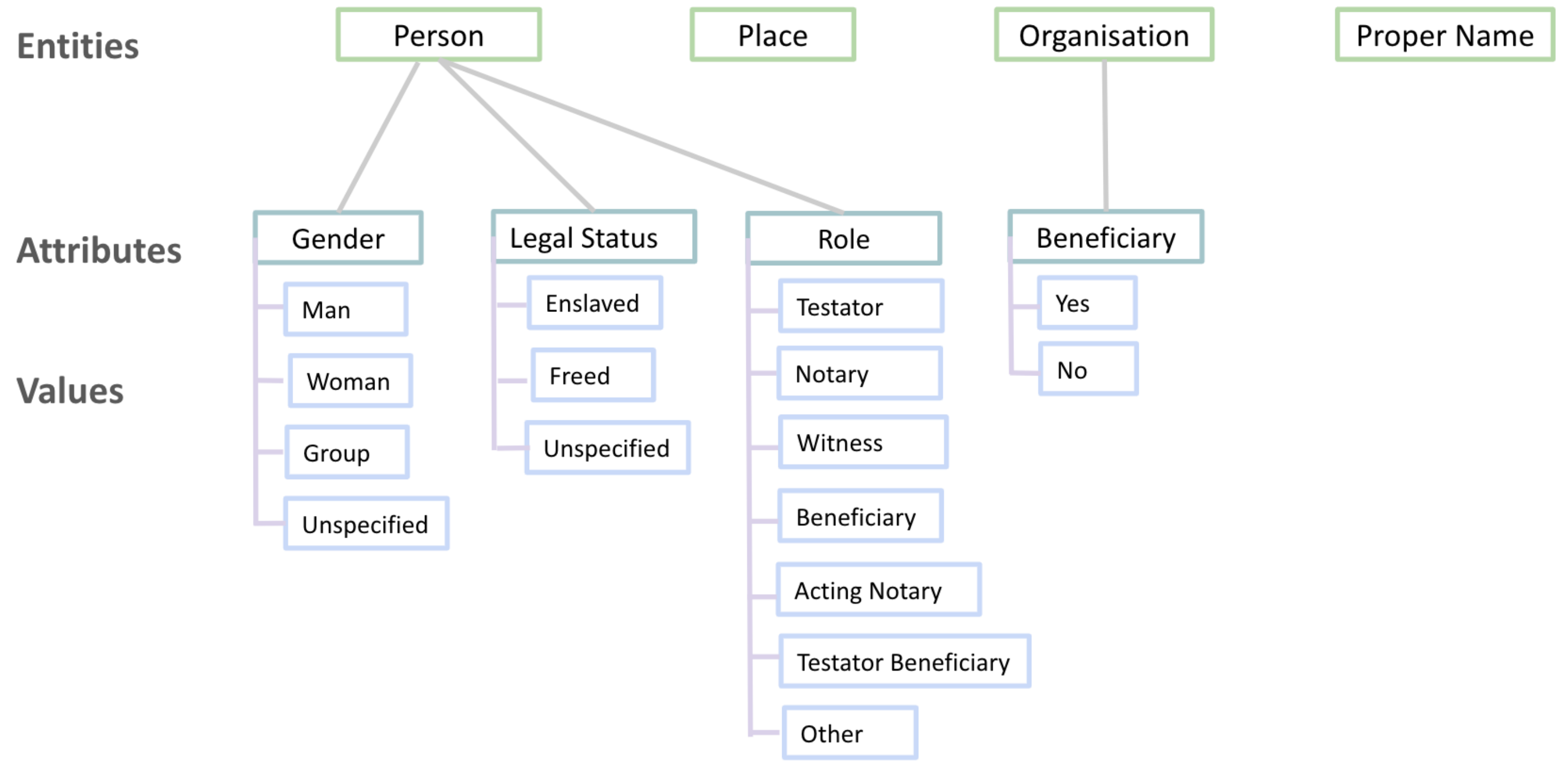}
    \caption{Proposed annotation typology.}
    \label{fig:typology}
\end{figure}

To `unsilence' colonial archives by broadening access, more inclusive finding aids are required, encompassing all persons mentioned in the archive with emphasis on marginalized ones. Existing generic typologies for named entity recognition and classification tasks such as CoNLL~\citep{tjong_kim_sang_introduction_2003} or ACE \citep{doddington2004automatic} mainly focus on the high-level `universal' or `ubiquitous' triad \textit{Person}, \textit{Organization} and \textit{Location}~\citep{ehrmann_named_2016}. These entity typologies alone are insufficient to overcome the challenges we face, for two main reasons. Firstly, they focus exclusively on \textit{named} entities, while colonial archives also contain traces of unnamed persons: we need to broaden the scope of the typology to incorporate mentions of unnamed persons too. Secondly, colonial archives, and the VOC testaments archive more specifically, are rich in further information about entities of interest, for example their role, gender, legal status. This information is also important in view of enriching finding aids, and can be captured via entity attributes. To address these needs, we propose a fit for purpose NER annotation typology. Our custom typology extends the universal triad to encompass all mentions of entities, both named and unnamed, and further qualifies them by gender, legal status, notarial roles and other relevant attributes. What is more, an initial exploratory pilot was conducted in order to acquaint the annotators with the corpus and task at hand, and to consolidate the annotation typology and guidelines. The final typology is illustrated in Figure~\ref{fig:typology}.

Typically, in named entity recognition and classification ``the word `Named' aims to restrict [Named Entities] to only those entities for which one or many rigid designators [..] stand for the referent''~\citep[130]{nadeau2007survey}. In our custom typology, only the entity type \textit{Proper name} corresponds to this definition of the named entity. It thus separates the name of an entity (always tagged separately as \textit{Proper Name}) from a generic reference to an entity type (\textit{Person}, \textit{Place} or \textit{Organization}). We introduce this distinction primarily because marginalized persons are frequently mentioned in the VOC testaments, and in colonial archives more generally, without name. Instead they are referred to by a variety of terms such as ``slaaf'' [slave], ``leiffeigenen'' [serf] and ``inlandse burger'' [indigenous person with some privileges]. \textit{As a result, what we propose is an Entity Recognition and Classification task, expanding on the scope of classic Named Entity Recognition and Classification.} The detection of entities, either named or unnamed, introduces a further degree of complexity in the task since it might make annotation guidelines less precise with respect to the exact boundaries of annotations.

In what follows, we present the annotation typology and mention related annotation guidelines, with examples from the corpus. 

\subsection{Person}\label{Person}

The entity type \textit{Person} may refer to individuals or groups of people. When annotating a text span as a person, the span should include the proper name and/or available contextual ``trigger words''~\citep{ehrmann_named_2021}. Trigger words in this typology also include words or phrases which provide information on the gender, legal status or notarial role of the person(s). Accordingly, the entity type person has three attributes: \textit{Gender}, \textit{Role} and \textit{Legal Status}. When a person is mentioned multiple times across a testament (with or without trigger words), they are annotated with the same attribute which was inferred from the presence of the trigger words. An example is provided in Figure~\ref{fig:legal status unspecified}.

\subsubsection{Gender}

\begin{figure}[h!]
    \centering
    \includegraphics[width=.17\linewidth]{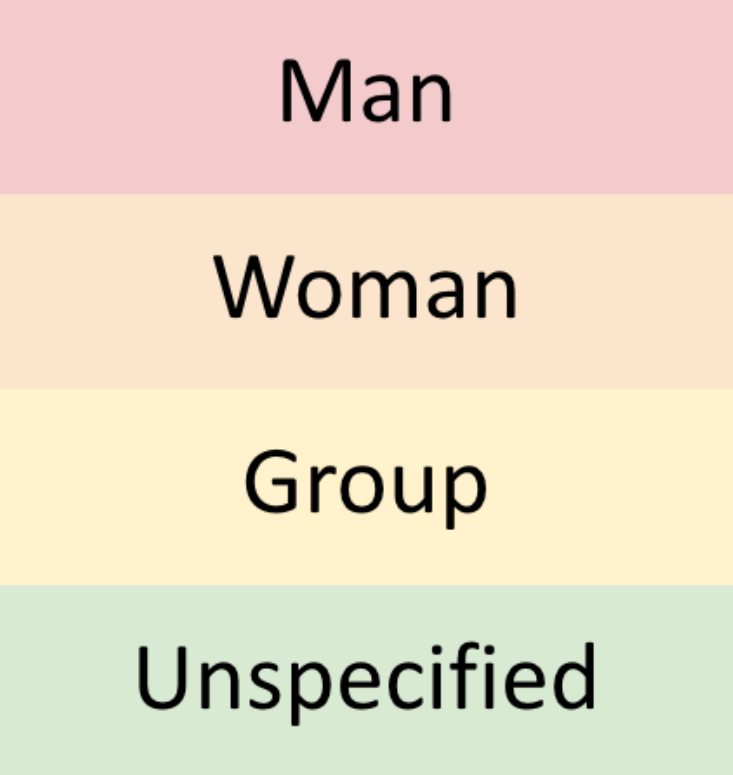}
    \caption{Legend for labeling person-gender attribute values.}
    \label{fig:person legend}
\end{figure}

\begin{figure}[h!]
    \centering
    \includegraphics[width=.75\linewidth]{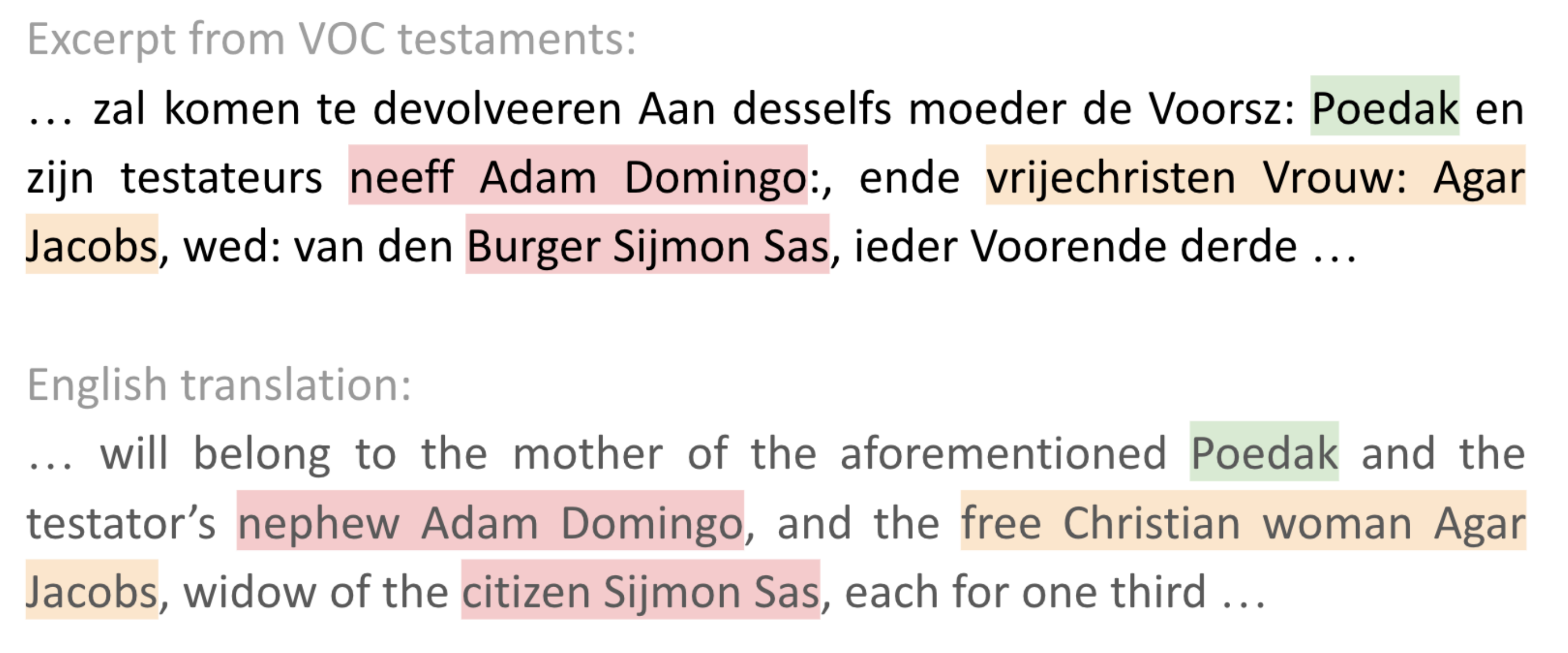}
    \caption{Instance of annotations of genders of persons, with and without leading qualifiers.}
    \label{fig:leading qualifier}
\end{figure}

\begin{figure}[h!]
    \centering
    \includegraphics[width=.77\linewidth]{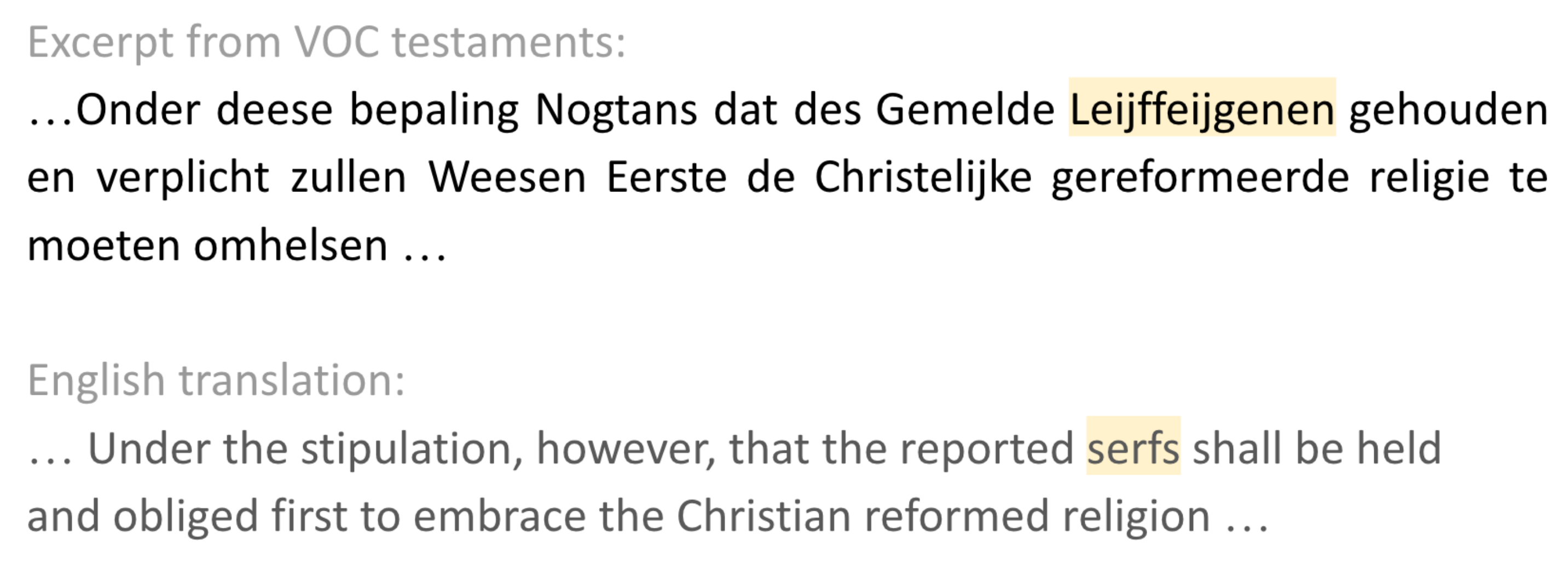}
    \caption{Instance of an annotation of a group of persons.}
    \label{fig:Gender group}
\end{figure}

\begin{figure}[h!]
    \centering
    \includegraphics[width=.73\linewidth]{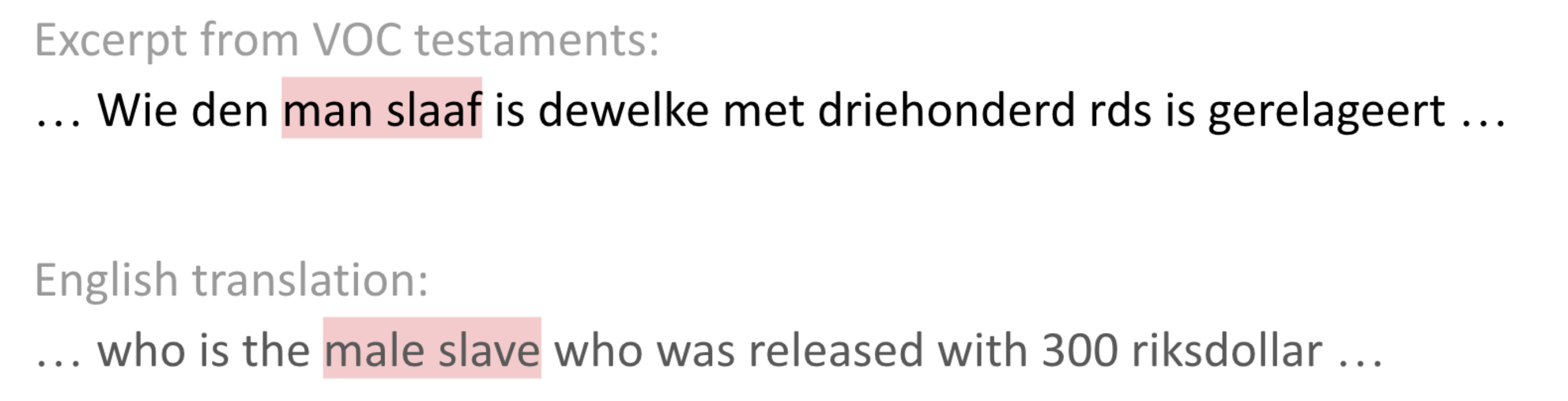}
    \caption{Instance of an annotation of a person mentioned without name.}
    \label{fig:no name person}
\end{figure}

When the mention of a person is followed or preceded by trigger words which reveal their gender, the text is annotated as a \textit{Person} with the appropriate value of the attribute \textit{Gender}. Figures \ref{fig:leading qualifier}, \ref{fig:Gender group} and \ref{fig:no name person} are examples of annotations for the gender attribute. For each entity person, the attribute gender takes exactly one of the values from the legend in Figure~\ref{fig:person legend}.

\noindent When a person is mentioned without a gender trigger word, their gender is marked as \textit{Unspecified}. This approach restricts possible `annotator bias' due to unfounded inferences. 

\noindent Persons are annotated by trigger words alone, in the absence of a proper name and in the case marginalised persons such as enslaved and formerly enslaved persons. This is because such persons are often mentioned without name and are of particular interest to our research. An example of a mention of an enslaved man without name is given in Figure~\ref{fig:no name person}.

\subsubsection{Legal Status}

\begin{figure}[h!]
    \centering
    \includegraphics[width=.16\linewidth]{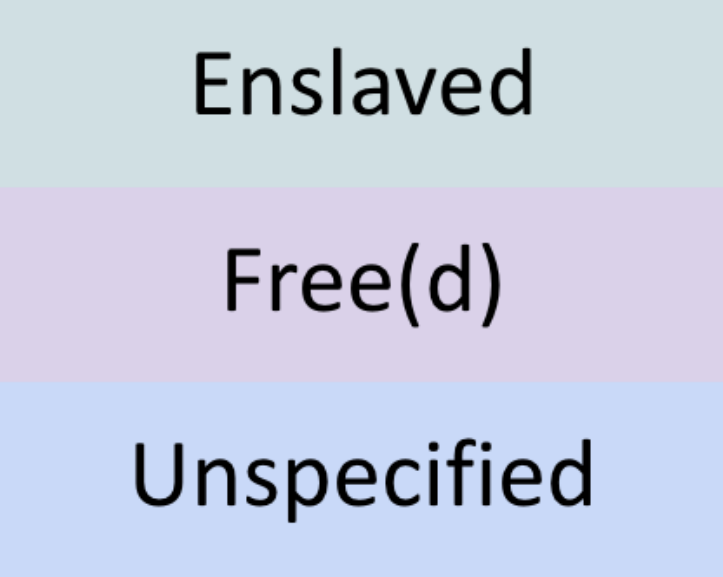}
    \caption{Legend for tagging person-legal status attribute values.}
    \label{fig:legendLegal}
\end{figure}

\begin{figure}[h!]
    \centering
    \includegraphics[width=.77\linewidth]{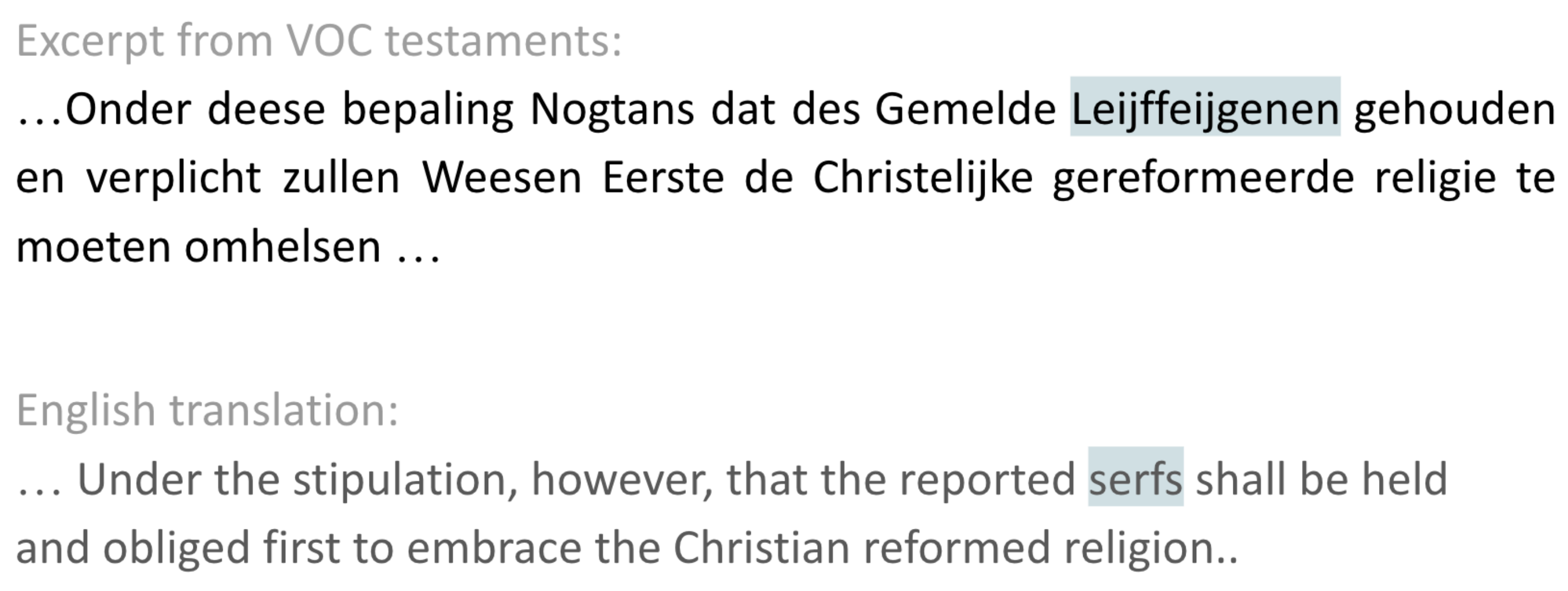}
    \caption{Instance of an annotation of persons with legal status of enslavement.}
    \label{fig:legal status enslaved}
\end{figure}

\begin{figure}[h!]
    \centering
    \includegraphics[width=.75\linewidth]{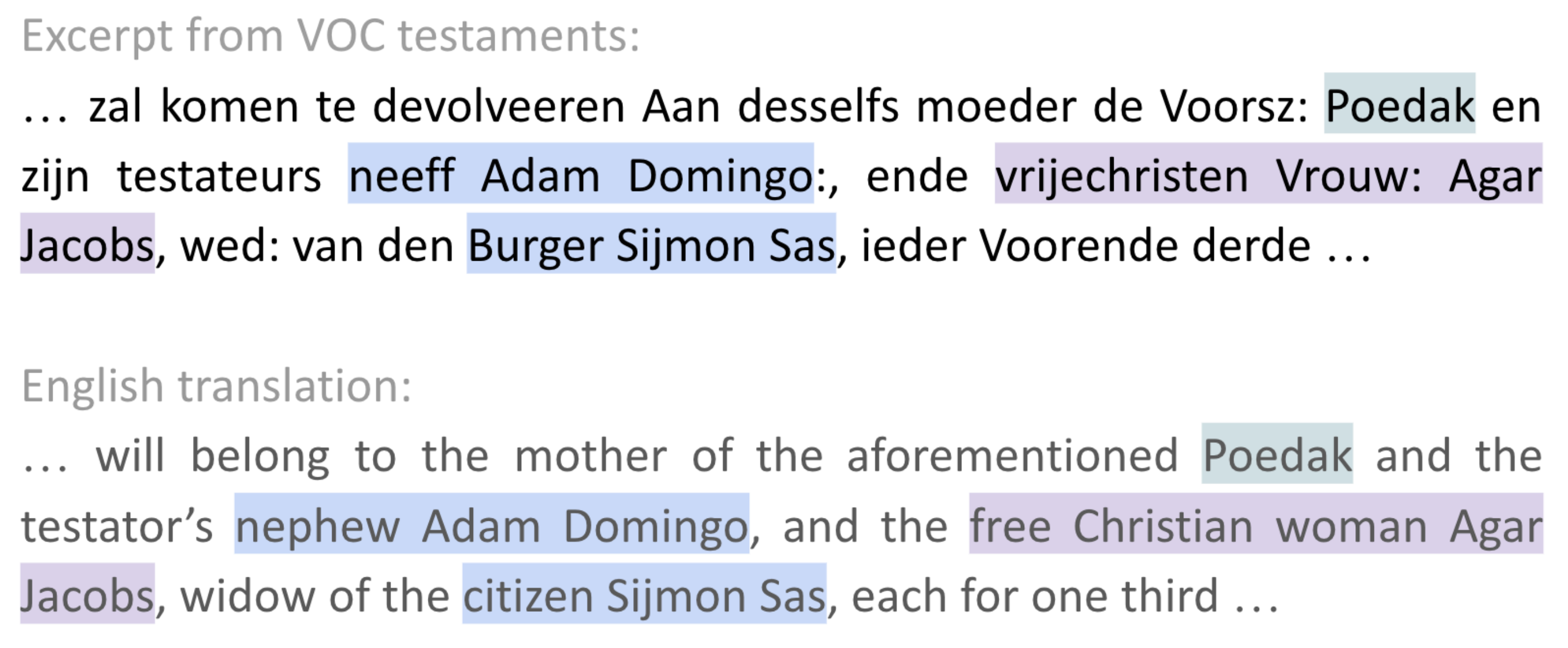}
    \caption{Instance of annotations of the legal status of persons.}
    \label{fig:legal status unspecified}
\end{figure}

For each entity \textit{Person}, the attribute \textit{Legal Status} takes exactly one of the values explained using the legend in Figure \ref{fig:legendLegal}. Figures~\ref{fig:legal status enslaved} and \ref{fig:legal status unspecified} contain examples of annotations for the legal status attribute.  

\noindent The attribute legal status takes the value \textit{Enslaved} when the trigger words clearly identify the individual(s) to be enslaved (see Figure \ref{fig:legal status enslaved}). The attribute value \textit{Free(d)} is often triggered by the word `vrije' [free]. It refers to persons who were set free (for different reasons such as when they bought themselves free, as an act of benevolence, or for economic reasons) sometimes on the condition that they adopted the Christian religion. It could also refer to children of the manumitted slaves who, although born free, kept carrying the adjective `vrije' [free], or if they were Christian they were labelled as `free Christian'. Finally, the adjective `free' was also used for groups of free indigenous (who were never enslaved) labelled for instance as `vrije inlander' [free native]. The attribute value \textit{Free(d)} captures these three different senses of the word `vrije', for which there is no clear way to clearly disambiguate among. When no trigger words are used, legal status is instead annotated as \textit{Unspecified} (see Figure \ref{fig:legal status unspecified}).

\noindent In the excerpt in Figure~\ref{fig:legal status unspecified}, ``Poedak'' has the attribute value \textit{Enslaved} in the absence of a trigger word that indicates enslavement. The reason is that Poedak is mentioned earlier (aforementioned) on the same testament as ``lijffeijgenen in name Poedak van Macassar'' [serf with name Poedak of Macassar]. Accordingly, all mentions of ``Poedak'' in the testament are labelled as a person with legal status enslaved.

\subsubsection{Role}

The attribute \textit{Role} refers to roles specific to testaments in notarial archives, which may take exactly one of the following values illustrated in Figure~\ref{fig:typology}: \textit{Testator}, \textit{Notary}, \textit{Witness}, \textit{Beneficiary}, \textit{Acting Notary}, \textit{Testator Beneficiary} or \textit{Other}.

\begin{figure}[h!]
    \centering
    \includegraphics[width=.73\linewidth]{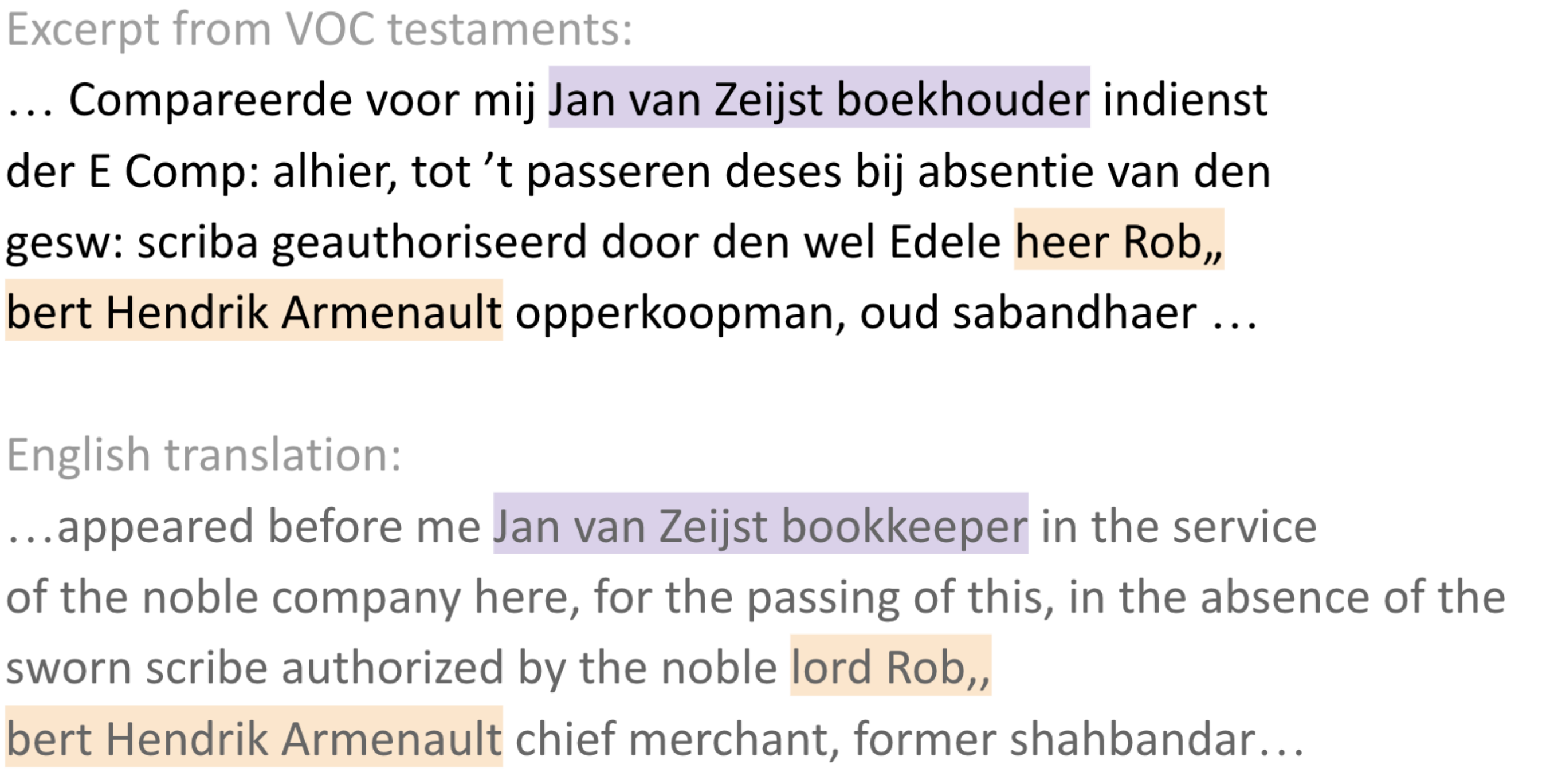}
    \caption{Instance of annotations of an acting notary and a person with role: other.}
    \label{fig:acting notary}
\end{figure}

An \textit{Acting Notary} is a role taken on by a person who, in the absence of an officially recognized notary, performs the notarial deed as can be inferred from the extract in Figure~\ref{fig:acting notary}. The role \textit{Testator Beneficiary} is attributed to those people who are both testator and beneficiary in the context of the testament. For instance, when man and wife collectively write down their testaments, each of them is a testator and often both of them are also each-other's beneficiaries. The role \textit{Other} is attributed to those persons whose role does not correspond to any of the six roles (for instance the annotation in orange in Figure~\ref{fig:acting notary}) or is when their role is not clearly mentioned.

\subsection{Place}

\begin{figure}[h!]
    \centering
    \includegraphics[width=.73\linewidth]{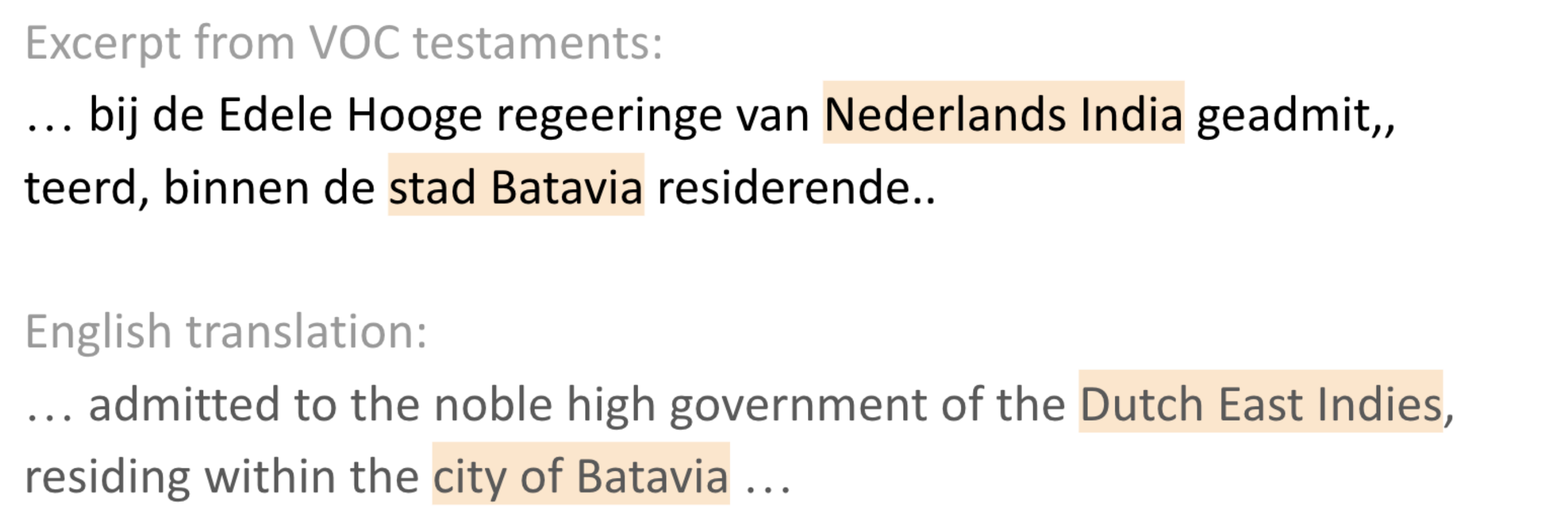}
    \caption{Instances of annotations of places.}
    \label{fig:place}
\end{figure}

The entity \textit{Place} is used to annotate places or locations. This entity is often called \textit{Location} in other typologies such as CoNLL~\citep{tjong_kim_sang_introduction_2003}. Consider Figure~\ref{fig:place} where place is annotated in yellow. 

\subsection{Organization}

\begin{figure}[h!]
    \centering
    \includegraphics[width=.75\linewidth]{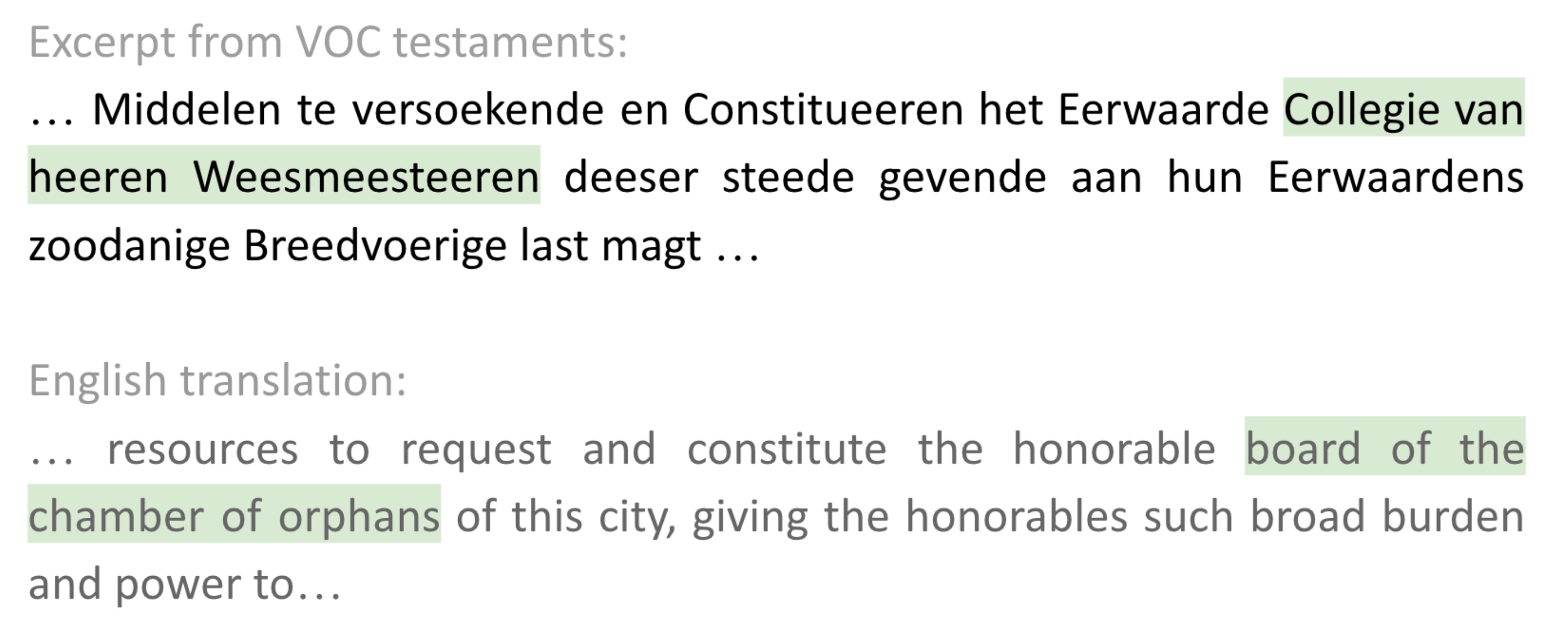}
    \caption{Instance of an annotation of an organization.}
    \label{fig:organization}
\end{figure}

This entity, as the name suggests, refers to organizations such as companies, orphanages, religious institutions and other branches of the church. Organizations have the attribute \textit{Beneficiary} which can take the value \textit{Yes} or \textit{No} depending on whether the testator decides an organization to be their beneficiary. Figure~\ref{fig:organization} is an instance of the latter.

\subsection{Proper Name}

\begin{figure}[h!]
    \centering
    \includegraphics[width=.73\linewidth]{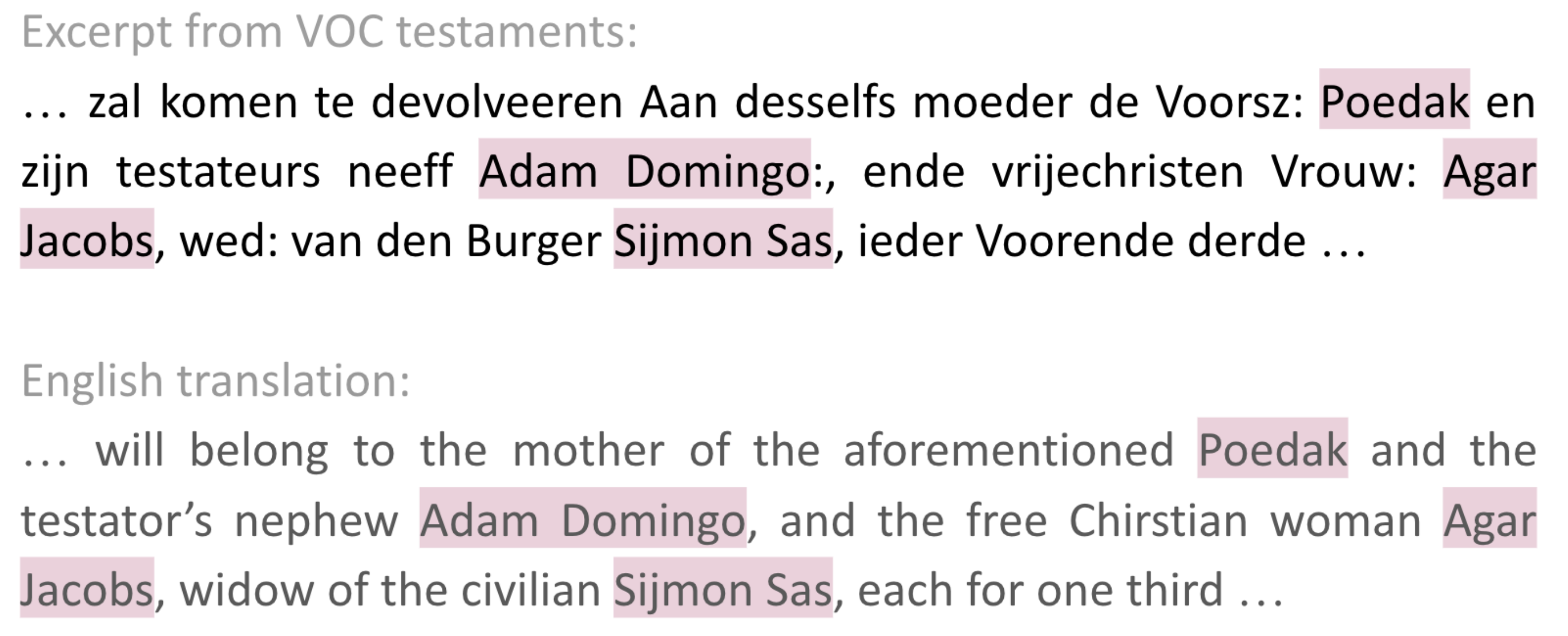}
    \caption{Instances of annotations of proper names.}
    \label{fig:propername}
\end{figure}

The entity \textit{Proper name} refers to names (proper nouns) of the other entities in this typology: \textit{Person}, \textit{Place} and \textit{Organization}. In Figure~\ref{fig:propername}, proper names are annotated in pink, which can be compared with Figure~\ref{fig:leading qualifier} and Figure~\ref{fig:legal status unspecified} where the same excerpt is labeled using the entity person and attributes gender and legal status respectively. In our dataset, annotations overlap.

\section{Results}

\subsection{Annotated Corpus}

\paragraph{HTR Quality} The VOC testament texts were extracted via Handwritten Text Recognition (HTR) by the Dutch National Archives, using a model combining ground truth from 17\textsuperscript{th} and 18\textsuperscript{th}-century VOC records. The ground truth for this combined model consists of 4810 manually transcribed pages from the VOC archives. The Dutch National Archives report an HTR Character Error Rate (CER) of 5.3 on a test set and 7.3 on a held-out sample set\footnote{\url{https://noord-hollandsarchief.nl/ontdekken/nhalab/project-transkribus-2}.}. The CER for a given page is calculated as:
    
$$\mathrm{{CER}} = \frac{i+s+d}{n}\times 100$$
 
where $n$ is the number of characters inclusive of spaces; $i$, $s$ and $d$ are the minimum number of insertions, substitutions and deletions respectively, required to attain the ground truth result from the HTR text. 

\begin{figure}[h!]
    \centering
    \includegraphics[width=0.5\textwidth]{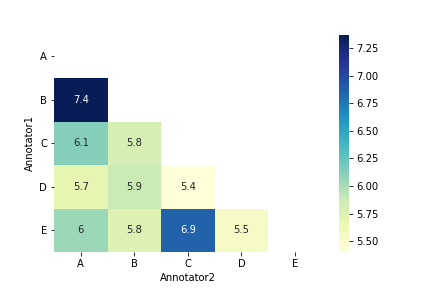}
    \caption{Overlap of annotated pages between each pair of annotators, calculated via Jaccard's distance. The overlapping pages are used to calculate the inter-annotator agreement.}
    \label{fig:annotedoverlaps}
\end{figure}

\paragraph{Corpus Selection} At the onset of the project, it was unclear by which criteria testaments are grouped into a bundle and also whether there is a logic in the order of the 51 extant bundles. Given this ambiguity and in the attempt to capture as much variation in content and transcription quality as possible, 13 non-consecutive and equally spaced (i.e., every fourth) bundles have been selected for annotation. From each of these, a range between 15-50\% pages have been annotated (as much as possible). Each bundle contains on average 1200 pages, thus 180-600 pages have been annotated per bundle. Each annotator was allocated a fixed number of pages per bundle. Furthermore, for each resulting bundle-annotator pair, 10\% of the pages have been duplicated into the sample allocated to two other annotators, in order to calculate their inter-annotator agreement. During the pilot, we established that this overlap was sufficient for the stable calculation of inter-annotator agreement. Given that different annotators advanced at different speeds and different pages contain varying amounts of text, the resulting overlap between each set of annotators might be less than 10\%, as shown in Figure~\ref{fig:annotedoverlaps}.

The Brat annotation tool~\citep{stenetorp2012brat} was used for manually annotating the corpus. While working on HTR texts in Brat, the annotators were invited to compared the texts with the scans provided on the website of the Dutch National Archives. This way of working proved instrumental in overcoming the limitations of HTR quality. The corpus is also provided in machine-readable IOB format (inside-outside-beginning); further details on the export from Brat to IOB are provided in the accompanying repository.

\paragraph{Corpus Characteristics} The corpus consists of 2193 unique pages, plus 307 duplicated ones for calculating the inter-annotator agreement, resulting in a total of 2500. This corresponds to roughly 4\% of the entire VOC testaments archive. The total number of annotations is 68,429, of which 32,203 at entity level (47\%) and 36,226 at attribute level (53\%); more details are given in Table~\ref{tab:annotations_entities}. The total number of annotated tokens is 79,797. We divide the corpus of annotations into three splits: training (70\%), validation (10\%), and test (20\%). We randomly sample annotated pages into splits by applying stratified sampling over annotation typologies and annotators, to maintain the overall data distribution over splits. 

\begin{table}[h!]
  \begin{center}
    \begin{tabular}{c|r|r} 
      \textbf{Entity type} & \textbf{Number of annotations} & \textbf{Percentage over total} \\
      % &  & $\%$ \\
      \hline
      Person & 11,715 & 36.3 \%\\
      Place & 4510 & 14.0 \%\\
      Organization & 1080 & 3.5 \%\\
      Proper name & 14,898 & 46.2 \%\\
      \hline
      \textbf{Total} & \textbf{33,203} & 100 \%
    \end{tabular}
    \caption{Number and share of annotations per entity type.}
    \label{tab:annotations_entities}
  \end{center}
\end{table}

\begin{figure}[h!]
    \centering
    \includegraphics[width=0.99\textwidth]{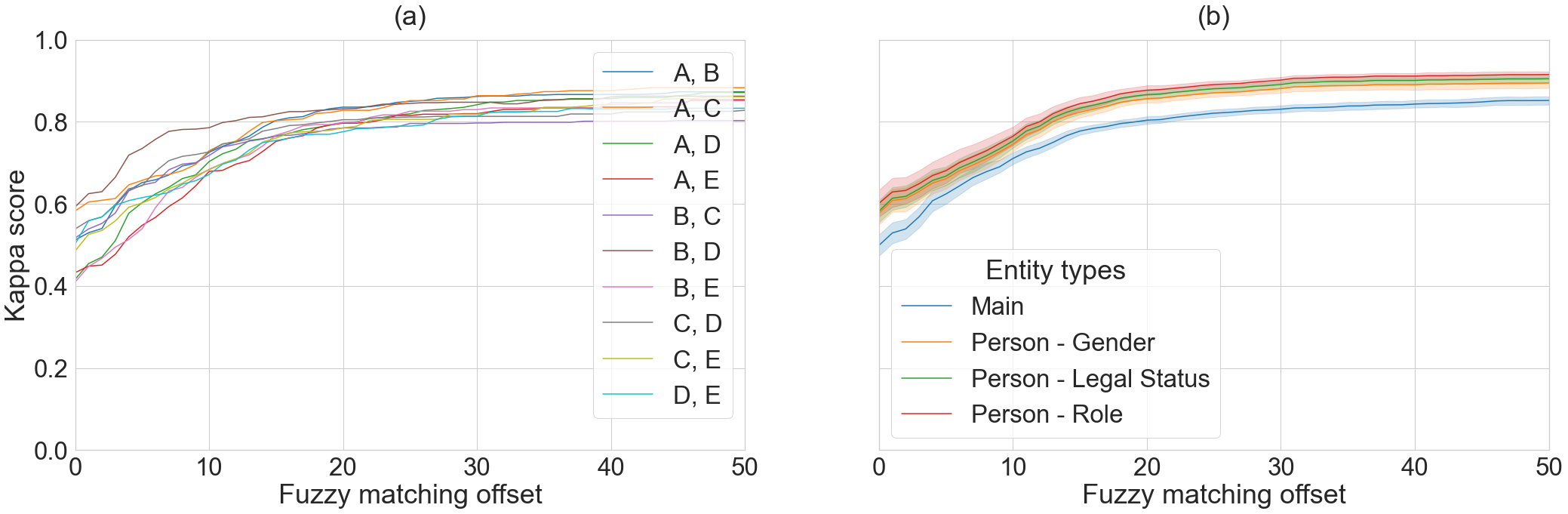}
    \caption{Inter-annotator agreement evaluation, considering  the Person, Place and Organization entities. (a) Scores between annotators. (b) Averages per entity, with 95\% bootstrapped confidence intervals.}
    \label{fig:iaa}
\end{figure}

\paragraph{Inter-Annotator Agreement} We use the Cohen's kappa score to evaluate the inter-annotator agreement~\citep{mchugh_interrater_2012}. We measure it both exactly and using a \emph{fuzzy matching offset}. This we define as the character offset that can exist between the same annotation given by two different annotators. Using an offset of 0 is equivalent to requiring an exact match, whereas an offset of 5 characters would entail considering two annotations to be the same if they overlap with a discrepancy of 5 characters at most. The inter-annotator agreement results between all pairs of annotators are shown in Figure~\ref{fig:iaa} (a), while the average scores per entity are in Figure~\ref{fig:iaa} (b). While with exact comparisons the kappa scores are only of moderate quality (0.5-0.6), with a modicum of fuzziness they converge to acceptable or strong values of 0.7-0.8 (at the 10 character offset mark).  

\subsection{Entity Recognition Baselines}
\label{ner}

In order to provide for a strong baseline to our proposed task, we make use of the best model configuration established in recent  work on NER for historical documents~\citep{todorov_transfer_2020-1}. This model text representation layer combines a variety of embeddings, including character-level embeddings and those produced from trained BERT models. We here use BERTje~\citep{vries_bertje_2019}, a state-of-the-art Dutch version of BERT. All embeddings are concatenated and followed by a Bi-LSTM-CRF layer~\citep{huang_bidirectional_2015}. We use and compare single-task and multi-task approaches. The former focuses on learning one entity type at a time, whereas the latter combines these tasks into a single model. Finally, we include results from the same baseline scorer used in the CLEF-HIPE-2020 challenge~\citep{arampatzis_overview_2020}: a Conditional Random Fields model~\citep{10.5555/645530.655813} based on the CFRsuite implementation~\citep{CRFsuite} and exposed via the \texttt{sklearn\_crfsuite} package.

We use the data splits established as described above, and execute three runs with different random seeds and identical model configuration, reporting results on the run achieving best results over the validation set. Our scoring approach follows best practices from CLEF-HIPE-2020~\citep{arampatzis_overview_2020,ehrmann_extended_2020}: all metrics are calculated as micro averages at annotation level (not the token level). The document-level macro average is given in two different flavors: as the average of micro scores across pages, and as the overall macro average at annotation level. We remind the reader that a document corresponds to a transcribed page in our corpus; in this section only, we use `document' in a machine learning sense. Furthermore, we use \emph{strict} and \emph{fuzzy} scoring. The former only considers exact boundary matching, whereas the latter includes overlapping boundaries. More specifically, fuzzy scoring considers predictions as correct if they have \emph{at least} one token overlap with the ground truth (and the predicted tag is the correct one). We report precision (P), recall (R) and F1 Score (F).

Entity-level results for \textit{Person}, \textit{Place} and \textit{Organization} (`Main entities') are shown in Table~\ref{tab:ner-results-main}. Micro average scores for the entity type \textit{Person name} and all attributes are shown in Table~\ref{tab:ner-results-person}, except those for \textit{Organization beneficiary} which are given separately in Table~\ref{tab:ner-results-org}. The best scores for each metric and tag are in bold. What emerges is that the task is clearly difficult with results remaining low in particular for recall, while faring better for precision. Some tags are more difficult to detect than others, specifically \textit{Organization beneficiary}. While a direct comparison is not possible, our results are not too distant from those achieved by~\cite{hendriks_recognising_2020} on similar archival records and focusing on the \textit{Person} entity type only. Lastly, we underline how the CRF baseline remains a strong option, even when compared with neural network-base approaches. The CRF baselines appears to be more precise, while the neural network architecture usually achieved better recall. This result is partially different from the conclusions drawn by the CLEF-HIPE-2020 challenge~\citep{arampatzis_overview_2020}, and warrants further study.

\begin{table}[h]
    \centering
    \begin{subtable}[h]{1\textwidth}
        \centering
        \begin{tabular}{c|cccccc|cccccc}
            \multirow{3}{*}{\textbf{Model}} & \multicolumn{6}{c|}{\textbf{Main}} & \multicolumn{6}{c}{\textbf{Main - Macro}} \\
            & \multicolumn{3}{c}{\textbf{Fuzzy}} & \multicolumn{3}{c|}{\textbf{Strict}} & \multicolumn{3}{c}{\textbf{Fuzzy}} & \multicolumn{3}{c}{\textbf{Strict}} \\
                &  P  &  R  &  F  &  P  &  R  &  F &  P  &  R  &  F  &  P  &  R  &  F \\
            \midrule
            \textbf{CRF baseline} & \textbf{.73} & .56 & \textbf{.63} & \textbf{.53} & .41 & \textbf{.46} & \textbf{.69} & .43 & .51 & .37 & .28 & .32\\
            \midrule
            \textbf{BERTje + Bi-LSTM-CRF} & .71 & .57 & \textbf{.63} & .51 & .41 & \textbf{.46} & .68 & \textbf{.53} & \textbf{.59} & \textbf{.47} & \textbf{.37} & \textbf{.41}\\
            \textbf{+ multi-task} & .69 & \textbf{.59} & \textbf{.63} & .49 & \textbf{.42} & .45 & .67 & \textbf{.53} & .58 & .45 & .35 & .39
        \end{tabular}
    \caption{Micro and macro averages.}
    \end{subtable}

    \begin{subtable}[h]{1\textwidth}
        \centering
        \begin{tabular}{c|cccccc}
            \multirow{3}{*}{\textbf{Model}} & \multicolumn{6}{c}{\textbf{Main - Macro, Document}} \\
            & \multicolumn{3}{c}{\textbf{Fuzzy}} & \multicolumn{3}{c}{\textbf{Strict}} \\
                &  P  &  R  &  F  &  P  &  R  &  F \\
            \midrule
            \textbf{CRF baseline} & \textbf{.73 $\pm$ .18} & .56 $\pm$ .22 & \textbf{.63 $\pm$ .19} & \textbf{.53 $\pm$ .28} & .41 $\pm$ .26 & \textbf{.46 $\pm$ .26}\\
            \midrule
            \textbf{BERTje + Bi-LSTM-CRF} & .71 $\pm$ .18 & .57 $\pm$ .21 & \textbf{.63 $\pm$ .18} & .51 $\pm$ .28 & \textbf{.42 $\pm$ .25} & .45 $\pm$ .25\\
            \textbf{+ multi-task} & .68 $\pm$ .19 & \textbf{.59 $\pm$ .22} & \textbf{.63 $\pm$ .18} & .48 $\pm$ .28 & \textbf{.42 $\pm$ .26} & .45 $\pm$ .26
        \end{tabular}
    \caption{Macro averages, document level aggregation.}
    \end{subtable}

   \caption{Results for the entity types Person, Place and Organization.}
   \label{tab:ner-results-main}
\end{table}

\begin{table}[h]
    \centering
    \begin{subtable}[h]{1\textwidth}
        \centering
        \begin{tabular}{c|cccccc|cccccc}
            \multirow{3}{*}{\textbf{Model}} & \multicolumn{6}{c|}{\textbf{Person name}} & \multicolumn{6}{c}{\textbf{Gender}} \\
            & \multicolumn{3}{c}{\textbf{Fuzzy}} & \multicolumn{3}{c|}{\textbf{Strict}} & \multicolumn{3}{c}{\textbf{Fuzzy}} & \multicolumn{3}{c}{\textbf{Strict}} \\
                &  P  &  R  &  F  &  P  &  R  &  F &  P  &  R  &  F  &  P  &  R  &  F \\
            \midrule
            \textbf{CRF baseline} & \textbf{.8} & .61 & \textbf{.69} & \textbf{.71} & .54 & \textbf{.61} & \textbf{.77} & .58 & .66 & \textbf{.47} & \textbf{.35} & \textbf{.4}\\
            \midrule
            \textbf{BERTje + Bi-LSTM-CRF} & .72 & \textbf{.65} & \textbf{.69} & .62 & \textbf{.56} & .59 & .73 & .59 & .65 & .37 & .3 & .33\\
            \textbf{+ multi-task} & .7 & .64 & .67 & .58 & .54 & .56 & .72 & \textbf{.64} & \textbf{.67} & .39 & \textbf{.35} & .37
        \end{tabular}
    \caption{Entity Person name and attribute Gender.}
    \end{subtable}

    \begin{subtable}[h]{1\textwidth}
        \centering
        \begin{tabular}{c|cccccc|cccccc}
            \multirow{3}{*}{\textbf{Model}} & \multicolumn{6}{c|}{\textbf{Legal status}} & \multicolumn{6}{c}{\textbf{Role}} \\
            & \multicolumn{3}{c}{\textbf{Fuzzy}} & \multicolumn{3}{c|}{\textbf{Strict}} & \multicolumn{3}{c}{\textbf{Fuzzy}} & \multicolumn{3}{c}{\textbf{Strict}} \\
                &  P  &  R  &  F  &  P  &  R  &  F &  P  &  R  &  F  &  P  &  R  &  F \\
            \midrule
            \textbf{CRF baseline} & \textbf{.77} & .6 & \textbf{.68} & \textbf{.58} & .45 & \textbf{.51} & \textbf{.76} & .53 & .63 & \textbf{.43} & \textbf{.3} & \textbf{.35}\\
            \midrule
            \textbf{BERTje + Bi-LSTM-CRF} & .7 & .63 & .66 & .5 & .44 & .47 & .71 & .6 & .65 & .33 & .28 & .3\\
            \textbf{+ multi-task} & .7 & \textbf{.66} & \textbf{.68} & .52 & \textbf{.48} & .5 & .71 & \textbf{.61} & \textbf{.66} & .33 & .28 & .3
        \end{tabular}
    \caption{Attributes Legal status and Role.}
    \end{subtable}

   \caption{Results for the entity type Person name and attributes; micro averages.}
   \label{tab:ner-results-person}
\end{table}

\begin{table}[h]
    \centering

        \begin{tabular}{c|cccccc}
            \multirow{3}{*}{\textbf{Model}} & \multicolumn{6}{c}{\textbf{Organization beneficiary}} \\
            & \multicolumn{3}{c}{\textbf{Fuzzy}} & \multicolumn{3}{c}{\textbf{Strict}} \\
                &  P  &  R  &  F  &  P  &  R  &  F \\
            \midrule
            \textbf{CRF baseline} & \textbf{.27} & .12 & \textbf{.17} & \textbf{.21} & \textbf{.09} & \textbf{.13}\\
            \midrule
            \textbf{BERTje + Bi-LSTM-CRF} & .07 & \textbf{.32} & .11 & .0 & .0 & .0\\
            \textbf{+ multi-task} & .26 & .11 & .15 & .16 & .07 & .1
        \end{tabular}
   \caption{Results for the attribute Organization beneficiary; micro averages.}
   \label{tab:ner-results-org}
\end{table}

\begin{figure}[h!]
    \centering
    \includegraphics[width=0.99\textwidth]{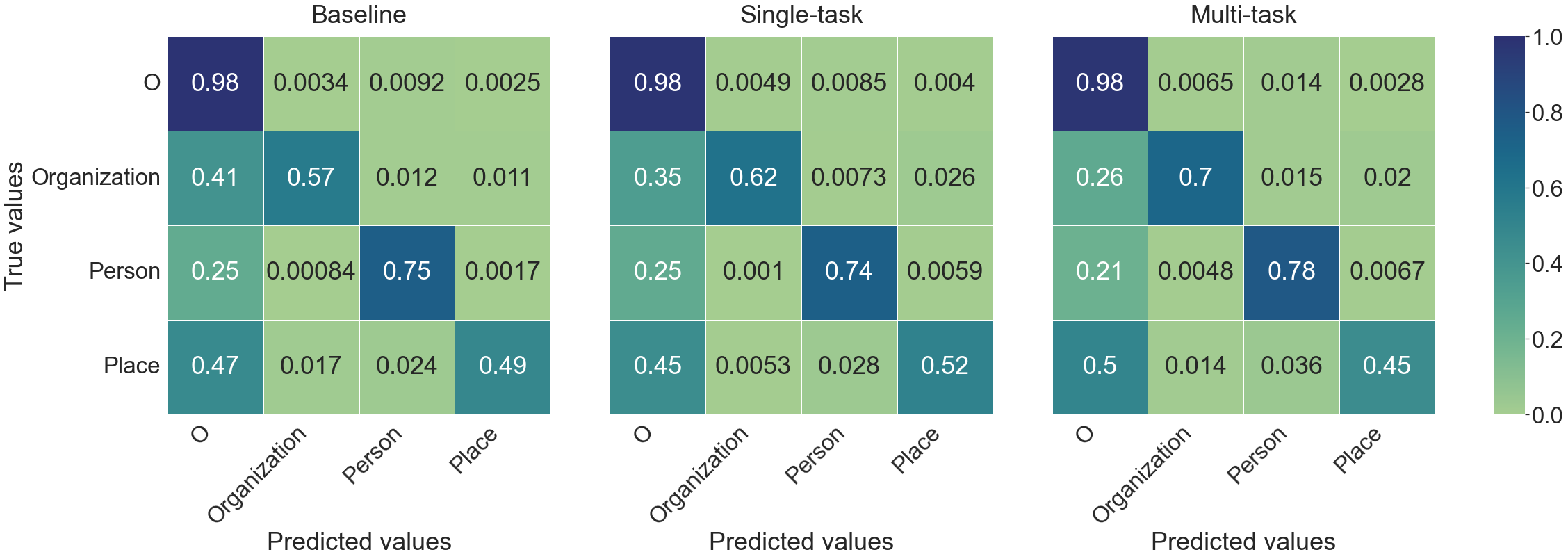}
    \caption{Normalized confusion matrices comparing ground truth and predictions for entities Person, Place and Organization.}
    \label{fig:cm_gt_predictions}
\end{figure}

Finally, in Figure~\ref{fig:cm_gt_predictions} we provide confusion matrices showing how predictions compare against the ground truth at the entity type level (\textit{Person}, \textit{Place}, \textit{Organization}). In general, we see how all models tend to over-predict the `O' tag (Outside of an annotation): a common issue in NER models, in part caused by the overabundance of `O' tokens. Mitigating this issue and thus improving recall, while also not loosing in terms of precision, constitutes a promising avenue for future work.

\section{Conclusion}

We started our work from the question of how it would be possible to find information on people that are hidden in archives. Extant indexes of historical records often embed systematic omission biases. The issue is particularly pressing for colonial archives, whose indexes ignore the presence of the enslaved and colonized. We have shown that it is possible to automatically and systematically surface information on marginalized groups even from these records. By considering the archives of testaments from the Dutch East India Company (VOC) as a case study, we have proposed an entity recognition shared task comprising an annotation typology, a corpus of annotations, and baseline results. Our main contribution entails expanding the task of named entity recognition to encompass mentions of unnamed historical entities, in particular persons. We found that, while challenging, the task appears doable even on such complex historical archival records. 

Nevertheless, the proposed approach to `unsilence' archival records might also be regarded as problematic: a form of disclosure leading to new, possibly dubious forms of categorization. Awareness of this issue has led us to avoid interpretations about which group people belonged to (e.g., by gender or legal status), unless this information was clearly stated in the sources. We also refrained from attempting to detect a persons' origin, since it would immediately result in problematic questions regarding ethnicity and its portrayal, in turn often erroneous and not helpful for modern-day finding aids. Finding the right balance for a respectful broadening of access to colonial records will require, going forward, a constant dialogue among archivists, scholars, governments, and the public.

We conclude by suggesting some directions for future work. Most immediately, the results from our baselines can be improved upon. After having reached satisfactory results, the proposed approach could be used to enrich the existing finding aids at the Dutch National Archives, turning this research into an application. In general, we deem important for more use cases from other colonial archives to be conducted, in such a way that more typologies, annotated corpora, models and other resources can be produced in the near future. These efforts should gradually be consolidated, primarily by devising a general typology and guidelines for entity recognition in colonial archives, providing for a shared conceptual ground all the while maintaining flexibility to accommodate the specificities of every archive. All efforts should not remain at the research stage but empower user-facing applications, in such a way that colonial archives can become more accessible and pluralized over time. To this end, the development of an ethical framework for the appropriate application of automation constitutes another key direction of future work.

\section*{Code and Data}

The corpus of annotations, codebase and the corresponding data card\footnote{The data card is based on and extends the accountability frameworks proposed by~\cite{gebru2021datasheets} (2021),~\cite{bender-friedman-2018-data} (2018) and~\cite{10.1145/3531146.3533231} (2022) to explicate assumptions and possible biases in dataset creation.} necessary to replicate our experiments can be found as a Zenodo archive: \url{https://doi.org/10.5281/zenodo.6958430}.

\section*{Acknowledgments}

We thank Saskia Virgina Noot, Thijs Vorstenburg and Clare Shutt for conducting the pilot\footnote{\url{https://www.nationaalarchief.nl/innovatie-in-archiefonderzoek-prijs}.} for this project. This work was made possible by the digitization efforts of the Dutch Nationaal Archief and we thank Milo van de Pol, Liesbeth Keijser and Diederick Kortlang for providing us with context on the testaments. Nadia F. Dwiandari from the Indonesian Arsip Nasional was helpful in providing some background information on the VOC-notary archives in Indonesia. We express our gratitude for the integral feedback of the participants of our workshop at The Critical Visitor\footnote{\url{https://www.universiteitleiden.nl/en/research/research-projects/humanities/critical-visitor}.} Field Lab which have been crucial in helping us in thinking about the politics of categories which led us to revise our typology. The dataset was made possible by the annotations created by researchers: Roos Bijleveld, Silja de Vilder Coombs, Emma Louise van der Hage, Jonas Guigonnat, Yolien Mulder and Bert van Splunter. Sincere thanks to Leon van Wissen for setting up the annotation software infrastructure and his insightful feedback at numerous points during this project. We thank our reviewers for their feedback. Finally, we thank the digital humanities research group CREATE\footnote{\url{https://www.create.humanities.uva.nl}.} at the University of Amsterdam and the Dutch Research Council (NWO, project number NWA.1228.192.108), for providing financial support. 

\bibliographystyle{unsrtnat}
\bibliography{references}  %%% Uncomment this line and comment out the ``thebibliography'' section below to use the external .bib file (using bibtex) .

\includepdf[pages=-,pagecommand={},width=1.15\textwidth]{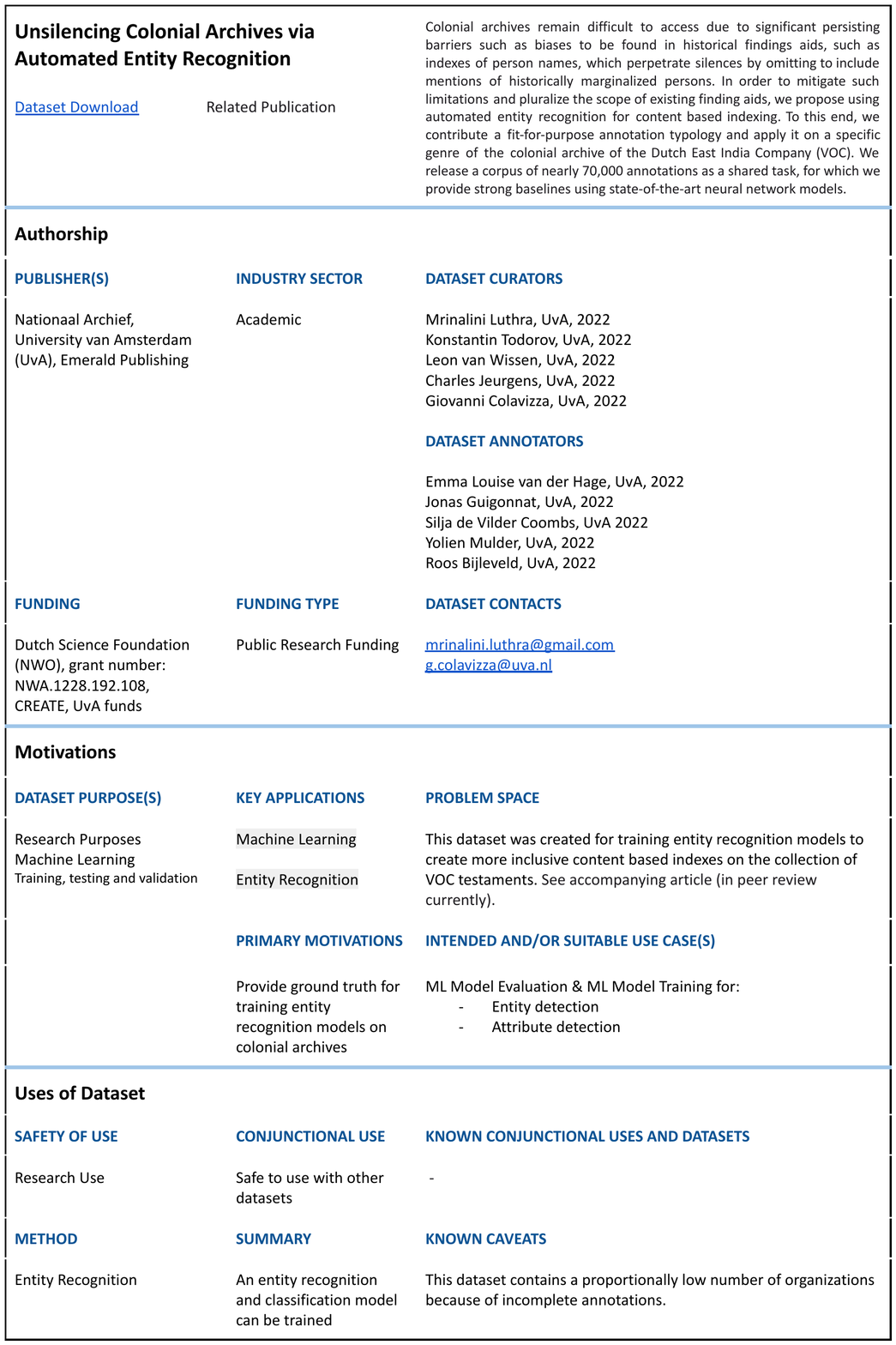}

\end{document}